\def \cm{~\rm{cm}}
\def \s{~\rm{s}}
\def \km{~\rm{km}}

\def \K{~\rm{K}}
\def \g{~\rm{g}}

\def \erg{~\rm{erg}}

\def \yr{~\rm{yr}}

\documentclass[12pt,preprint]{aastex}
\usepackage{natbib}

\shortauthors{Soker}

\begin{document}

\title{APPLYING THE JET FEEDBACK MECHANISM TO CORE-COLLAPSE SUPERNOVA EXPLOSIONS}

\author{Noam Soker\altaffilmark{1}}

\altaffiltext{1}{Dept. of Physics, Technion, Haifa 32000, Israel;
soker@physics.technion.ac.il.}

\begin{abstract}
I examine a mechanism by which two fast narrow jets launched by a newly formed
neutron star (NS), or a black hole (BH), at the center of a core collapse supernovae (CCSN),
form two slow massive wide (SMW) jets.
Such SMW jets are assumed as initial conditions in some numerical simulations that
demonstrate that SMW jets can expel the rest of the collapsing star.
The original fast narrow jets must deposit their energy inside the star via shock waves,
and form two hot bubbles that accelerate a much larger mass to form SMW jets.
To prevent the jets from penetrating through the still infalling gas and escape
instead of forming the hot bubbles, the jets should encounter fresh infalling gas.
This condition is met if the jets' axis changes its direction.
The exact condition is derived.
In addition, to maintain a small neutrino cooling the fast narrow jets must be
shocked at a distance $r \ga 10^3 \km$ from the core, such that most of the
post-shock energy is in radiation, and temperature is not too high.
The scenario proposed here was shown to be able to suppress star formation
in newly formed galaxies, and in forming SMW jets in cooling flow clusters of galaxies
and in planetary nebulae.
Namely, I suggest that NSs (or BHs) at the center of CCSNs shut off their own growth and
expel the rest of the mass available for accretion by the same mechanism that super-massive BHs
shut off their own growth, as well as that of their host bulge, in young galaxies.

\end{abstract}


\section{INTRODUCTION}
\label{sec:intro}

Observations and theoretical considerations support the existence of
slow massive wide (SMW) jets (outflows) in variety of astronomical objects.
SMW jets have the following properties.
(1) Slow: Their outflow velocity is much lower than the
   escape velocity from the powering compact object.
(2) Massive: Their mass outflow rate is about equal or larger that the
   accretion rate onto the compact object.
(3) Wide: Each of the two {{{ high Mach number }}} jets covers a solid angle of $ \ga 1$.
Namely, it has a half-opening angle of $\alpha \ga 30^\circ$.
{{{ The limit of $\sim 30^\circ$ is found in numerical simulations of jets in
planetary nebulae (Akashi \& Soker 2008) and in clusters of galaxies (Sternberg \& Soker 2008).
In the next section it is noted that initially narrow jets that have a large fraction of
thermal energy, i.e., a relatively small Mach number, behave like wide jets because the
thermal pressure causes them to expand transversely. }}}

SMW jets are powered by accretion, most commonly through an accretion disk,
of mass onto a compact objects.
In most cases the SMW jets are not launched directly neither by the accretion disk
nor by the compact object; they are rather secondary outflows.
{{{ In the case of core collapse supernovae (CCSNe) the SMW jets might be
powered by neutrinos, or the narrow jets that form the SMW jets might be powered
by neutrinos. The energy of course is accretion energy. }}}
The most striking example is presented in the seminal work of Moe et al.
(2009 ApJ submitted).
By conducting a thorough analysis, Moe et al. (2009) find the outflow
from the quasar SDSS J0838+2955 to have a velocity of $\sim 5000 \km \s^{-1}$,
and a mass outflow rate of $\sim 600 M_\odot \yr^{-1}$,
assuming a cover fraction of $\delta \simeq 0.2$.
Earlier observations also suggest that AGN can have SMW jets
(de Kool et al. 2001; Crenshaw \& Kraemer 2007; Behar et al. 2003; Kaspi \& Behar 2006).
On the theoretical side, it has been shown that SMW jets can inflate the
bubbles observed in cooling flow clusters of galaxies
(Sternberg et al. 2007; Sternberg \& Soker 2008).
Wide jets formed by white dwarfs (WD) accreting mass from asymptotic giant
branch stars can shape some planetary nebulae (Soker 2008, henceforth Paper 1).

In all these cases the SMW jets are not disk winds.
The power in these outflows is large, and an accretion close to the compact object
is required.
The following mechanism to form SMW jets was studied in Paper 1.
The accretion disk launches two opposite narrow jets with velocity of the order of the
escape velocity from the accreting object and with a mass outflow rate of
$\sim 1-20 \%$ of the accretion rate, as in most popular models for fast narrow jet launching.
However, the fast narrow jets encounter surrounding gas that originates in the mass accretion
process (hence the name `feedback'), and are terminated by strong shocks.
Two hot bubbles are formed from the post-shock fast jets' material.
These bubbles accelerate the surrounding gas to form two SMW jets
that are more massive and much slower than the fast narrow original jets.
There are two conditions for this mechanism to work.
(1) {{{ In many cases }}}
the jets must encounter new segments of the surrounding gas, such that they don't
have the time to `drill' a hole and penetrate through the surrounding gas.
{{{ (If the jets are wide enough and/or the surrounding gas dense and thick enough,
then this condition is not required. In section \ref{subsec:pen} I show that in the present
study this condition is indeed required.) }}}
This can be achieved when the surrounding gas and the compact object have a
relative velocity to each other.
Either there is an ordered velocity (like in a binary orbital motion),
or the surrounding gas is turbulent.
Else, the jets might precess rapidly enough (Paper 1).
(2) The radiative energy losses by the post-shock gas must be small.

In paper 1 it was shown that the two conditions can be met for the fast jets blown
by AGN in cooling flow clusters and by WD accreting at a high rate.
These conditions can also be met during the epoch of galaxy formation (Soker 2009,
henceforth Paper 2).
When the narrow jets launched by the super massive black hole (SMBH) at the center of
the newly formed galaxy do not penetrate the infalling gas, they are able
to expel the infalling gas and by that to prevent further star formation.
The condition for the jets not to penetrate the infalling gas lead to a relation
between the SMBH mass and the mass of the galaxy (or the bulge; Paper 2).
This relation is similar to the observed relation.

In the present paper I examine the conditions for narrow jets that are
launched by an accretion disk around the newly formed neutron star (NS; or a BH)
in CCSNe, to expel the infalling gas and turn the
infall to an explosion.
In section 2 I briefly review the role of jets in CCSNe.
In Section 3 I derive the conditions for the jets not to cool by neutrino loses
(sec.  3.1), and not to penetrate and escape the surrounding gas (sec. 3.2).
I discuss the results and their implications in section 4, and summarize in
section 5.

\section{JETS IN CORE-COLLAPSE SUPERNOVAE}
\label{sec:CCSN}

Jets can play a key role in exploding CCSNe (e.g., Khokhlov et al. 1999;
MacFadyen et al. 2001; Maeda \& Nomoto 2003; Woosley \& Janka 2005; Couch et al. 2009).
Khokhlov et al. (1999) injected a jet with a radius of $r_{j0}=1200 \km$ at a distance of
$R_{\rm in}=3820\km$ from the center.
The mass in the two jets was $\sim 0.1 M_\odot$, and their speed $\sim 0.1 c$.
Practically they injected SMW jets.
As the jets' launched by the newly formed NS are likely to be narrow and fast
(a velocity of $>0.1 c$), the formation of the jets launched
by Khokhlov et al. (1999) should be explained.

Couch et al. (2009) found that to match observations their jets must start with
a large fraction of thermal energy (their model v1m12).
The total mass in the jets in their model v1m12 was $0.12 M_\odot$, and the maximum velocity
$10^4 \km \s^{-1}$.
As evident from their fig. 7, such a jet practically starts as a wide jet.
Namely, their initial conditions for their model v1m12 was that of a SMW outflow.
As Sternberg et al. (2007) have shown, wide jets are very efficient in driving a large
outflow of the surrounding gas.
Sternberg et al. (2007) simulated SMW jets in clusters of galaxies, but I emphasize
that the physics is same, e.g., as in planetary nebulae and the great eruption of Eta Carinae
(Paper 1).
MacFadyen et al. (2001) injected jets at $R_{\rm in}=50 \km$, but their jets were injected at
a much later time in the explosion, and are less relevant to the present paper.
In any case, they also showed that SMW jets are efficient in removing
the envelope further out.
When narrow jets are simulated (e.g., Alloy et al. 2000; Zhang et al. 2003, 2004)
no envelope ejection occurs.
Indeed, narrow jets that maintain a constant direction do not expect to expel the
non-turbulent surrounding gas (Sternberg et al. 2007).
In another set of simulations, Zhang et al. (2006) added a SMW outflow to
their simulations of relativistic jets.
The papers cited above further emphasize the need to answer the question as of
how such SMW jets can be formed in CCSNe.

In the present Paper I argue that SMW jets, as simulated by, e.g., Khokhlov et al. (1999) and
Couch et al. (2009), can be formed when a fast ($v_j > 0.3 c$) and light
(total mass of $ \sim 0.01 M_\odot$) jets interact with the infalling
gas at distances of ($10^3 < r < 10^4 \km$) to the center.
I will follow the same procedure that was used to explain the formation of SMW jets in
other astrophysical systems (Paper 1), and also gave a theoretical correlation
between the masses of SMBHs and their host bulges that is similar to the
observed one (Paper 2).

I do note that MacFadyen \& Woosley  (1999) find a disk-wind in their simulations,
that has the same properties as SMW jets.
However, I emphasize again that a disk wind blown from an extended disk surface
is less efficient {{{ (energetically speaking) }}} than SMW jets formed from
shocked narrow jets that are launched from the very inner region of the accretion disk.
{{{ The reason is that a wind from an extended disk region is launched from
a shallow potential well. In the proposed model, the SMW jets are powered by narrow jets
that are launched from the inner disk, where the potential well is much deeper, and the accreted
gas released much more gravitational energy. }}}
Indeed, MacFadyen \& Woosley  (1999) simulate the formation of a BH (and not a NS), and
the strong disk wind is formed only when the collapsing core is rapidly rotating.
Kohri et al. (2005) conducted a study of disk wind in CCSN, where the central object is a NS.
{{{ They propose that the wind energy is able to revive a stalled shock and
help to produce a successful supernova explosion. }}}
Here again, the wind comes from an extended region in the disk, and it is less efficient
that the expected fast jets blown from the very inner region of the disk.
More over, to form their proposed disk wind, Kohri et al. (2005) require the progenitor's
core to rotate very rapidly, as they form the accretion disk earlier than in the present model.

\section{THE JET-FEEDBACK MECHANISM}
\label{sec:jet}

I start with a list of assumptions that has a one to one correspondence with the
the list of assumptions that I used in paper 2 for the explanation of
SMBH-bulge masses correlation in galaxies.
Different are the typical values used, as here I deal with a CCSN.
Like, e.g., Couch et al. (2009), I assume that about a Chandrasekhar mass has
already collapsed and formed the almost final neutron star, with an escape velocity of
$v_{\rm esc} \simeq 0.5c$;
I don't deal with the first stage of the collapse that forms the compact object.
\begin{enumerate}
\item The feedback mechanism, i.e., the one that expels infalling mass from close regions
to large distances is driven by jets launched by an accretion disk around the newly formed
NS (or BH).
{{{ In the present study the infalling gas at the late stage of disk formation falls
from a distance of $r_s \sim 3000 \km$, and the condition for the formation of an accretion disk
at $r_{\rm NS} \sim 15 km$ is that it rotate at $ \ga (r_{NS}/r_s)^{1/2} \simeq 0.07$
of break-up velocity at $r_s$.
This is plausible (Ott et al. 2006), but not clear yet if it is common enough.
Observations of radio pulsars imply that the initial rotation period of NSs is tens to
hundreds of milliseconds (Kaspi \& Helfand 2002),
which is $\sim 0.01$ of the break-up velocity of NSs.
This shows that the accreted gas has indeed a relative large amount of angular momentum.
Newly born main sequence stars rotate at $\sim 1-10 \%$ of their break-up velocity
(Rodriguez-Ledesma et al. 2008), presumably because of their interaction with the accretion disk.
Therefore, the same process can explain the rotation of NSs at birth.
However, whether accretion disks can be formed in most CCSN events,
is a subject of a future research. In particular, the possibility that stochastic accretion
of dense clumps instead of a smooth accretion should be examined with 3D numerical codes.
Such an accretion can lead to a temporarily formation of an accretion disk,
with rapid changes of axis direction (see below).
Over all, I estimate that the condition for the formation of an accretion disk is that
the material at the edge of the pre-collapse core rotates at $\ga 0.01$ times its
break-up velocity.  }}}
\item The properties of jets launched by NS (or BH) have some universal properties,
such as the fast jet's speed $v_f \simeq v_{\rm esc} \simeq 0.5c$.
\item There is a universal ratio between mass lose rate in the two jets to mass accretion rate.
Using the same ratio of ejected to accreted energy as in Paper 2, $0.05$, the
ejected to accreted mass ratio is $\eta \equiv \dot M_f/\dot M_{\rm acc} \simeq 0.1$.
\item The mass available for an inflow is very large.
Namely, the mass that is eventually accreted to the NS or BH is limited by the
feedback mechanism and not by the mass available in the core of the collapsing star.
\item In addition to the radial motion, there is a relative transverse (not radial) motion
between the newly formed compact object and the inflowing mass.
This transverse velocity can result from turbulence motion in the inflowing gas,
or from rapid changes in the jets' axis, e.g., precession or stochastic angular momentum
direction of the accreted gas.
The maximum possible transverse velocity in the infalling gas is of the order
 of the free fall velocity of the inflowing gas.
According to numerical simulation (e.g, Burrows et al. 2007; Dessart et ak. 2006; Ott et al. 2009),
in CCSN without fast rotation the infalling gas is shocked at a distance
of $R_s \sim 50-200 \km$ from the center, and the post-shock region becomes convective
with transverse speeds of $\sim 5 \times 10^4 \km \s^{-1}$.
However, as is explained in section \ref{subsec:cooling}, the jets are required to
be shocked at much larger distances, where the free fall velocity is only $\sim 10^4 \km \s^{-1}$.
For the change in the jets' axis direction I assume that the jets' axis changes
its direction over a time scale about equal to the last phase of accretion when
the jets are formed, $\sim 0.1 \s$.
This is $\sim 100$ times the Keplerian period on the surface of a NS.
A change in one radian over 0.1 second at a distance of $r_s \sim 3000 \km$ results
in a relative transverse velocity of $v_{\rm rel} \simeq 3 \times 10^4 \km \s^{-1}$.
Note that the transverse velocity is not a matter velocity in this case, but rather
a change in jet's axis.
Over all, I take $10^4 \la v_{\rm rel} \la 5 \times 10^4 \km \s^{-1}$
when the jets are shocked at $r_s \sim 3000 \km$.
\item The surrounding mass $M_s$ that resides at a typical distance $r_s$ and
having a density $\rho_s$, is flowing inward at about the free fall velocity.
Thus, the mass inflow rate is $\dot M_s \simeq 4 \pi r_s^2 v_s \rho_s$.
This assumption should be better constrained by future 3D numerical simulations.
I will use that expression to derive the density of the surrounding medium in order to emphasize
the dynamical nature of the process, as in Paper 2.
I note that for an accretion rate of $\sim 1 M_\odot \s^{-1}$ at a radius od $r_s=3000 \km$,
with a free fall velocity at that radius ($v_s=1.1 \times 10^4 \km \s^{-1}$), the density is
$\rho_s \sim 10^6 \g \cm^{-3}$, very similar to that in the initial model used by Couch et al. (2009).
{{{ Note that the infall rate can be larger than the accretion rate, as the
narrow jets might expel part of the infalling gas, hence preventing it from being
accreted. This is the basic process of the feedback. }}}
\end{enumerate}

With these assumptions in hand, I proceed to describe the interaction of the jets with
the surrounding inflowing gas, as was done in Paper 2.
If the jets penetrate through the surrounding gas they will be collimated by that gas,
and two narrow collimated fast jets will be formed, similar to the flow structure in
the simulations of Sutherland \& Bicknell (2007) for AGN jets, and
of MacFadyen et al. (2001) for CCSN.
If, on the other hand, the jets cannot penetrate the surrounding gas they will deposit
their energy in the inner region.
Two hot bubbles (that might merge to one almost spherical bubble) will be formed that
will accelerate the surrounding gas and form a SMW outflow, as was studied in Paper 1.
The conditions for the jets not to cool are derived in section \ref{subsec:cooling}, while
the conditions for the jets not to penetrate the surrounding gas but rather form a SMW outflow
are derived in section \ref{subsec:pen}.

\subsection{NEUTRINO COOLING}
\label{subsec:cooling}

Based on the results of Itoh et al. (1989, 1996), Kohri et al. (2005),  and the approximate
equation given by MacFadyen \& Woosley  (1999), I approximate the neutrino cooling rate by
\begin{equation}
\epsilon_\nu \simeq 10^{25} \left( \frac{T}{10^{10} \K} \right)^9
\erg \cm^{-3} \s^{-1}.
\label{eq:cool1}
\end{equation}

I will consider first the possibility that the narrow jets are shocked within the
convective region at $r \sim 50-200 \km$ found in the simulations of Ott et al. (2009).
Taking a mass of $\sim 0.1 M_\odot$ to be shocked inside a radius of $<100 \km$,
and reside inside two bubbles that occupy part of the spherical volume,
$V \sim 10^{21} \cm^3$, the density is $\rho_b \sim  10^{11} \g \cm^{-3}$.
The total energy carried by neutrinos in a time $\Delta t$ is
\begin{equation}
E_\nu \simeq 10^{46}
\left( \frac{T}{10^{10} \K} \right)^9
\left( \frac{V}{10^{21} \cm^{-3}} \right)  \Delta t \erg,
\label{eq:cool2}
\end{equation}
where $\Delta t$ is given in seconds.
The infalling gas is accelerated to a velocity of $\sim 10^4 \km \s^{-1}$
and the relevant interaction radius is $\sim 10^3 \km$.
Therefore, $\Delta t \sim 0.1 \s$.
{{{ For the neutrino cooling }}} to be negligible compared with the jets'
energy $\sim 10^{51} \erg$, the temperature
of the shocked jets should be $T \la 4 \times 10^{10} \K$.
This limits the pre-shock jets' velocity (assuming the material was already disintegrated
to nucleons)  to $v_f \la 5 \times 10^4 \km \s^{-1}$.
This velocity is too low for the mechanism proposed here.
{{{ The jets' anergy of $\sim 10^{51} \erg$ is not derived from the model.
It is basically assumed. However, it does fit the process where only the last $\sim10 \%$ of
the accreted mass has sufficient specific angular momentum to form an
accretion disk, and $\sim 10 \%$ of the accretion energy in the
disk is converted to jets' kinetic energy. }}}

I turn to consider the possibility that the jets penetrate the small mass at $r \la 10^3 \km$, and
are shocked at a larger radius $r > 10^3 \km$.
Khokhlov et al. (1999) and Couch et al. (2009), for example, injected jets at
$R_{\rm in}=3800 \km$ from the center.
I take a mass of $\sim 0.01 M_\odot$ to be shocked at a radius of $r_s \simeq 3000 \km$,
and the two bubbles to occupy most of the volume inside $r_s$, $V \simeq 10^{26} \cm^3$.
In such a large volume the radiation energy $a T^4 V$ in the post shock region must be considered.
The temperature of the post-shock gas is
\begin{equation}
T \sim 7 \times 10^9
\left( \frac {v_f}{0.5 c} \right)^{1/2}
\left( \frac {M_f}{0.01 M_\odot} \right)^{1/4}
\left( \frac{V}{10^{26} \cm^{-3}} \right)^{-1/4}   K,     \quad {\rm for }  \quad V \ga 10^{24} \cm^{-3}
\label{eq:trad1}
\end{equation}
Using this expression for the temperature in equation (\ref{eq:cool2}), scaling with the
kinetic energy of the fast jets $E_f=(1/2)M_f v_f^2$, and using the distance of the
shock $r_s \simeq (0.25V)^{1/3}$, I find that the total energy carried by neutrinos
in a time $\Delta t$ in this volume is
\begin{equation}
E_\nu \simeq 3 \times 10^{49}
\left( \frac {E_f}{2 \times 10^{51} \erg} \right)^{9/4}
\left( \frac{r_s}{3000 \km} \right)^{-15/4}  \Delta t \erg.
\label{eq:cool3}
\end{equation}

Equation (\ref{eq:cool3}) shows that for neutrino loses to be negligible, the
narrow fast jets should be shocked at a distance of $r_s \ga 3000 \km$.
The formation of jets as used by Khokhlov et al. (1999) and Couch et al. (2009)
can be explained by this mechanism.
In the initial (pre-explosion) model used by Couch et al. (2009) the mass inside
$3800 \km$ is $\sim 1.6 M_\odot$. This is the mass assumed to collapse and form the
core that forms the NS, and is not treated by Couch et al. (2009).
The collapse time of this region is $\sim 1 \s$.

\subsection{THE NON-PENETRATION CONDITION}
\label{subsec:pen}

Let the fast jets from the inner disk zone have a mass outflow rate in both directions
of $\dot M_f$, a velocity $v_f$, and let the two jets cover a solid angle of
$4 \pi \delta$ (on both sides of the disk together).
The density of the outflow at radius $r$ is
\begin{equation}
\rho_f =  \frac {\dot M_f}{4 \pi \delta r^2 v_f}.
\label{eq:rhof}
\end{equation}
The jets encounter the surrounding gas residing within a distance
$r_s$ and having a typical density $\rho_s$.
The head of each jet proceeds at a speed $v_h$ given by the balance
{{{ of the pressure exerted by the shocked jet's material with that
of the shocked surrounding gas. }}}
Assuming supersonic motion this equality reads
$\rho_s v_h^2 = \rho_f (v_f-v_h)^2$, which can be solved for $v_h$
\begin{eqnarray}
\frac {v_f}{v_h}-1 =
\left( \frac {4 \pi \delta r_s^2 v_f \rho_s}{\dot M_f} \right)^{1/2}
\simeq \left( \frac {\delta \dot M_s v_f }{\dot M_f v_s} \right)^{1/2}
\nonumber \\
=12.2
\left( \frac {\dot M_s/\dot M_f}{100} \right)^{1/2}
\left( \frac {\delta}{0.1} \right)^{1/2}
\left( \frac {v_f}{0.5c} \right)^{1/2}
\left( \frac {v_s}{10^4 \km \s^{-1}} \right)^{-1/2},
\label{eq:vh1}
\end{eqnarray}
where in the second equality the mass inflow rate $\dot M_s \simeq 4 \pi \rho_s v_s r_s^2$
(by assumption 6), has been substituted.
The ratio ${\dot M_s/\dot M_f} \sim {100}$ is taken as it is assumed that $\sim 1 M_\odot$
is falling at radius $\sim r_s$, while $\sim 0.1 M_\odot$ is accreted at the last stage of
the NS formation, out of which $\sim 0.01 M_\odot$ is blown in the jets.
{{{ Most, or even all, of the $\sim 1 M_\odot$ falling gas at $r_s$ is expelled by the hot bubbles
formed by the shocked narrow jets gas, and turns into the SMW jets.   }}}
The time required for the jets to cross the surrounding gas and break out of it is given by
\begin{equation}
t_p \simeq \frac {r_s}{v_h} \simeq
\frac{r_s}{v_f}
\left( \frac {\delta \dot M_s v_f }{\dot M_f v_s} \right)^{1/2}
=0.3  
\left( \frac {r_s}{3000 \km} \right) \s ,
\label{eq:tp1}
\end{equation}
where in the last equality the same values as in equation (\ref{eq:vh1}) have been used.

{{{ If the jet maintains its direction relative to the surrounding gas
(namely, neither the jet's axis changes its direction nor the surrounding gas has a
transverse motion), then the jet's head moves through the surrounding gas with a velocity
given by equation (\ref{eq:vh1}).
As was shown for some other astrophysical objects analytically (Paper 1, 2) and
numerically (Sternberg et al. 2007), and as will be shown below for the present case,
the penetration of the jets through the surrounding gas is quite rapid when there
is no change of direction. Most of the energy in the jets then, is deposited
at large distances, far from the region where the energy is needed to expel falling gas.
However, if there is a relative transverse velocity, the jet's head must start its
penetration outward from a new place relative to the surrounding gas.
If the transverse velocity is fast enough, then before the jets's head breaks out
from the surrounding gas, the jets has moved to interact with new parts of the
surrounding gas.
This process prevents the jet from escaping the inner region,
and the energy is deposited where it is need. }}}
For an efficient deposition of energy to the inflowing gas, we require that there will
be a relative transverse motion between the jets and the inflowing gas, such that the jets
continuously encounter fresh mass.
The relevant time is the time that the relative transverse motion of the jet and surrounding
gas crosses the jet's width $\tau_s \equiv D_j/v_{\rm rel} $.
The width of the jet at a distance $r_s$ from its source is $D_j=2 r_s \sin \alpha$,
where $\alpha$ is the half opening angle of the jet.
For a narrow jet  $\sin \alpha \simeq \alpha \simeq (2 \delta)^{1/2}$, and
\begin{equation}
\tau_s = \frac {2 (2 \delta)^{1/2} r_s}{v_{\rm rel}}.
\label{eq:taus}
\end{equation}

The demand for efficient energy deposition, $\tau_s \la t_p$, reads then
\begin{equation}
\frac {\dot M_s}{\dot M_f} \ga
8   \frac {v_f v_s}{v_{\rm rel}^2}.
\label{eq:ms1}
\end{equation}
The number 8 comes from the geometry of a narrow jet with a relative transverse velocity to
that of the ambient gas. Namely, the jet geometry and the transverse relative motion
are crucial to understand the ejection of the SN envelope.

Using assumption 3 that $\dot M_f=\eta \dot M_{\rm acc}$ in equation (\ref{eq:ms1}),
and substituting typical values, e.g., from assumption 5, gives
\begin{equation}
\frac {\dot M_s}{\dot M_{\rm acc}} \ga
8 \eta
\frac {v_f v_s }{v_{\rm rel}^2} = 1.3
\left( \frac {\eta}{0.1} \right)
\left( \frac {v_f}{0.5 c} \right)
\left( \frac {v_s}{10^4 \km \s^{-1}} \right)
\left( \frac {v_{\rm rel}}{3 \times 10^4 \km \s^{-1}} \right)^{-2} .
\label{eq:mscluster}
\end{equation}
Note that although $v_f$ and $\eta$ appear as two parameters, actually there is one
parameter $p_m \equiv \eta v_f$ in the above condition.
Its physical meaning is that of the momentum of the material ejected in the original
fast jets per unit accreted mass to the NS.
The two other physical variables, ${v_{\rm rel}}$ and $v_s$ are taken from typical
expected velocities in the collapsing star.

\section{DISCUSSION}
\label{sec:discussion}

The implication of equation (\ref{eq:mscluster}) is as follows.
If during the final stage of the NS formation jets are formed, then if the still
infalling gas has an inflow rate which is more than about the accretion
rate (a factor 1.3 by eq. \ref{eq:mscluster}) at that time,
then the jets will not penetrate trough the infalling gas, but rather be shocked
and deposit their energy inside the region of the infalling gas.
{{{ It should be noted that the infall mass is the gas flowing inward at large
(hundreds to thousands of km) distances, while the accreted mass is the gas added
to the central NS or BH.
As most of the infalling mass at late stages is expelled by the narrow jets, the ifall rate can
be much larger than the accretion rate at late stages. }}}
For example, if the jets are blown during the accretion of the last $\sim 0.1 M_\odot$ into the
NS, then if more than $\sim 0.1 M_\odot$ are falling at $r_s \sim 3000 \km$ at that time, they
will be expelled by the jet.
As a comparison, for the feedback mechanism in galaxy formation the ratio in the analogue
to equation (\ref{eq:mscluster}) is $\sim 600$ instead of $\sim 1$.
Two points should be kept in mind here.
Firstly, what matters is the density of the surrounding gas at $r_s$.
It was related to the mass infall rate at $r_s$ through assumption 6 in section \ref{sec:jet}.
Secondly, there is a large uncertainty in the exact value of the numerical
coefficient in equation (\ref{eq:mscluster}).
However, even an uncertainty by a factor of $\sim 10$ allows us to arrive at the conclusions.

There is one more crucial difference between accretion in AGN and young stellar objects
on the one side, and the case studied here on the other side.
In the first two cases the accretion phase lasts for a times much longer than the
dynamical time at the inner radius of the accretion disc;
the ratio is $>5$ orders of magnitude.
For example, in young stellar objects (YSOs) the dynamical time (about one Keplerian orbit)
in the inner region of the disk is $\sim$day,
while the accretion disks might last for thousands of years or more.
In AGN the dynamical time in the inner part of the disk is hours to days
(depending on the SMBH mass), while disks are observed for tens of years and are
though to exist for much longer times.
In the core collapse scenario the relevant accretion time is $\sim 1 \s$,
{{{ which is the typical time of the collapse of material from thousands of km. }}}
The accretion disk will not form immediately, but rather is expected to be formed in the
last stage, {{{ when the accreted mass comes from large radii and has high enough specific
angular momentum to form a disk. }}}
This implies a relevant time scale that might be $\sim 0.1 \s$.
The Keplerian period at the surface of the NS is $\sim 0.001 \s$, a ratio of only $\sim 100$.
This implies that the accretion disk might not have time to completely relax.
For example, the direction of the angular momentum of the disk will fluctuate
during the accretion process.
{{{ These fluctuations might be strong as the accreted mass in the last stage passes
through a shock that posses large eddies and large scales departure from sphericity,
even if there is no initial angular momentum  (Ott et al. 2009). }}}
In addition, the magnetic axes of many NS are known to be inclined to the spin.
Over all, the jet direction is not expected to be constant, unless the core of the
collapsing star has large angular momentum (see below).
Here I assumed that the jets' axis changes its direction (one radian) over the
relevant accretion phase of $\sim 0.1 \s$.

But even if the value of $v_{\rm rel}$ is lower
{{{ than what was assumed in its scaling in equation (\ref{eq:mscluster}), }}} and
it is equal to the free fall velocity of $\sim 10^4 \km \s^{-1}$ at $r_s \simeq 3000 \km$
(the transverse velocity in the infalling gas results from instabilities and shocks),
the fast narrow jet will still be arrested by the infalling mass.
The fast jets are shocked and form two hot bubbles
(that might merged to one bubble).
The hot bubbles accelerate the infalling gas over a very large solid angle.
The wide outflow has more mass than the originally narrow jets, and by
energy conservation the wide outflow has a lower velocity.
A slow massive wide (SMW) outflow (jets) has been formed.
It is this SMW outflow that will expel the rest of the stellar gas.
That a wide outflow can expel the stellar gas has been studied analytically by
Kohri et al. (2005), and was demonstrated in numerical simulations
by Couch et al. (2009), where their high thermal energy jets are practically SMW jets
(see section \ref{sec:CCSN} here).
The results here offer an explanation to the formation
of the jets simulated by Couch et al. (2009) in their v1m12 model.

The processes described above will be modified if the collapsing
core has a large angular momentum.
{{{ At present it is hard to calculate the required specific angular momentum for
this modification in behavior.
It probably requires the interaction with a binary companion close to the core, most
likely a companion that experienced a common envelope phase.
A crude estimate gives that the original core's mass at $r_s \sim 3000 \km$
should rotate at $\ga 0.1$ times its break-up velocity for a strong modification
of the process. For the formation of an accretion disk
with varying axis direction my (crude) estimate for the rotation velocity was
$\ga 0.01$ times its break-up velocity. }}}
The high specific angular momentum implies that the NS (or BH) spin axis and the
accretion disk axis directions are the same, and the accretion disk axis maintains
its direction as accretion proceeds.
Also, rapidly rotating cores do not form a strong convective region when
their gas falls and shocked (Ott et al. 2009).
Over all, the relative transverse velocity $v_{\rm rel}$ decreases.
Because the coefficient in equation (\ref{eq:mscluster}) is $\sim 1$, and the
expression is sensitive to $v_{\rm rel}$, a decrease in
$v_{\rm rel}$ by an order of magnitude might alow the original fast narrow jets
to penetrate the envelope.
The jets will not form a large enough hot bubbles, the SMW jets will be weaker,
and not all the stellar envelope will be expelled.
As a consequence more mass will be accreted leading to the formation of a BH.
A high accretion rate to the BH will lead to the formation of highly relativistic
jets, as required for the formation of gamma ray bursts (GRBs).
The accretion of more mass onto a BH might results in much more energetic
jets, up to a total energy of $>10^{52} \erg$, as observed in some GRBs (Cenko et al. 2009).

The suggestion that rapidly rotating cores lead to the formation of GRBs was raised before from
different considerations (e.g., Woosley \& Heger 2006).
Woosley \& Bloom (2006) emphasize that the suggestion that most rapidly rotating
and most massive stars form GRBs is a conjecture still to be proven.
Here I suggest a mechanism by which only rapidly rotating cores allow jets to
escape when they collapse.

\section{SUMMARY}
\label{sec:summary}

I examined a mechanism by which jets launched by the newly formed NS (or BH)
at the center of a core collapse SN (CCSN) expel the rest of the stellar mass.
More specifically, the mechanism converts fast narrow jets blown by the accreting newly formed
NS, to slow massive wide (SMW) jets, similar to those simulated by, e.g., Couch et al. (2009).
The basic assumption is that after a compact NS is formed, the final accretion
stage results in the formation of an accretion disk that launches two fast narrow jets.
Basically, the same process as occurs in the formation of young stellar objects.
{{{ The presence of jets in GRBs shows that jets can be formed in CCSN.
 However, it does not tell us the frequency of occurrence of jets. }}}
Six additional assumptions, that are listed at the beginning of section \ref{sec:jet},
are supplied.
These six assumptions are the same as those used in building the feedback model to
explain the correlation between the mass of supermassive black holes (SMBH)
and the mass of their host bulges (Paper 2).
All these assumptions are plausible and supported by different observations
and theoretical arguments.

The basic process to allow the jets to deposit their energy into the infalling mass
and turn the infall to an outflow, requires that the jets do not
penetrate the surrounding infalling gas; if they do, they escape the star as
two collimated fast jets. In addition, non-adiabatic cooling should be small;
in the present case cooling by neutrinos should be small.
For not penetrating the infalling gas the jets should encounter new material before
they escape.
Namely, the typical time for the jet's axis to cross the jet's width,
or for the infalling gas to cross the jet's width by transverse motion, at radius $r_s$,
should be shorter than the penetration time at radius $r_s$, $\tau_s \la t_p$.
This leads to equation (\ref{eq:mscluster}).
Due to precession of the newly formed NS magnetic axis and
a stochastic mass supply to the disk, the jets' axis is expected to
change its direction.
This change of jets' axis forces the jets to encounter different regions of the infalling gas.

If the mass inflowing rate $\dot M_s$ is larger than the value given by
equation (\ref{eq:mscluster}) the jets are very efficient in depositing
their energy to the inflowing gas.
As we consider the accretion of the last $\sim 0.1 M_\odot$, by condition (\ref{eq:mscluster})
it is enough that $0.05-0.5M_\odot$ are continue to fall at $r_s \sim 3000 \km$
that the jets will not escape.
This condition is quite easily met by CCSNe. Hence, we expect newly formed NS to be
able to explode the star.
The original jets are shocked and form two hot bubbles (that might merge),
with a total energy content of $\sim 10^{51} \erg$.
These bubbles accelerate the infalling gas outward to form a more massive, wider,
and slower outflow.
These SMW jets (outflow) have about the same energy as the original fast narrow jets,
and can therefore expel the rest of the envelope (Couch et al. 2009).
By the assumptions used here, the jets contain only $\sim 10 \%$ of the energy
available in the accretion of the last $\sim 0.1 M_\odot$.
If a NS is formed, the rest of the energy is mainly in neutrinos.

To keep neutrino loses negligible, the jets must be shocked at a distance of
$ r _s \ga 3000 \km $ from the center (section \ref{subsec:cooling}).
The hot bubbles occupy a large enough volume such that most of the jets'
kinetic energy is transferred to radiation energy.
The bubbles' temperature is low enough for neutrino losses to be small.

As evident from equation (\ref{eq:mscluster}) the condition is sensitive to the value
of $v_{\rm rel}$.
A rapidly rotating pre-collapse core is likely to lead to a more stable jets' axis.
This reduces the value of $v_{\rm rel}$, and we might reach a point where
condition (\ref{eq:mscluster}) is violated.
The original jets will escape the star, such as in GRBs, and will be less
efficient in expelling the infalling gas.
More gas will be accreted, and a BH is likely to be formed.
More energy will be released as accretion continues, eventually expelling the
rest of the stellar envelope.
Hence a SN might be formed after all.

Future studies will have to find whether the explosion is driven by SMW outflows
formed directly by the accretion disk (MacFadyen \& Woosley 1999; Kohri et al. 2005), or
whether SMW jets that are formed at larger distances are more important
(Khokhlov et al. 1999; Couch et al. 2009).
If jets are formed at large distances from the core, then the present paper
offers an explanation for their formation.

The scenario proposed here is in a preliminary stage.
It has a  big advantage in that the same basic mechanism can work to suppress star
formation in newly formed galaxies (Paper 2) and in forming wide jets
in cooling flow clusters and planetary nebulae (Paper 1).
Namely, I suggest that NSs (or BHs) at the center of CCSNs shut off their own growth
and expel the rest of the mass available for accretion by the same mechanism that
SMBH shut off their own growth, as well as that of their host bulge, in young galaxies.

\acknowledgements
I thank Eli Livne and Milos Milosavljevic, for useful comments,
{{{ and an anonymous referee for many detail comments that improved and clarified
the paper. }}}
This research was supported by the Asher Fund for Space
Research at the Technion, and by the Israel Science Foundation.

\end{document}